\documentclass[showpacs,aps,prl,twocolumn,superscriptaddress]{revtex4-1}
\usepackage[english]{babel}
\usepackage{graphicx}
\usepackage{color}
\usepackage{xcolor}
\usepackage{amsmath}
\usepackage{amssymb}
\usepackage[pdftex,breaklinks]{hyperref}

\newcommand{\Figref}[1]{Fig.~\ref{#1}}
\newcommand{\Eqref}[1]{Eq.~(\ref{#1})}

\newcommand{\revision}[1]{\textcolor{black}{#1}}

\newcommand{\rem}[1]{}

\begin{document}

\title{Single-photon emission mediated by single-electron tunneling \mbox{in plasmonic nanojunctions}}

\author{Q.~Schaeverbeke}
\affiliation{Univ.~Bordeaux, CNRS, LOMA, UMR 5798, F-33405 Talence, France}
\affiliation{Donostia International Physics Center (DIPC), E-20018, Donostia-San Sebasti\'an, Spain}
\author{R.~Avriller}
\affiliation{Univ.~Bordeaux, CNRS, LOMA, UMR 5798, F-33405 Talence, France}
\author{T.~Frederiksen}
\affiliation{Donostia International Physics Center (DIPC), E-20018, Donostia-San Sebasti\'an, Spain}
\affiliation{Ikerbasque, Basque Foundation for Science, E-48013, Bilbao, Spain}
\author{F.~Pistolesi}
\affiliation{Univ.~Bordeaux, CNRS, LOMA, UMR 5798, F-33405 Talence, France}

\begin{abstract}
Recent scanning \revision{tunneling} microscopy (STM) 
experiments reported single-molecule fluorescence induced by tunneling currents 
in the nanoplasmonic cavity formed by the STM tip and the substrate.
The electric field of the cavity mode couples with the current-induced
charge fluctuations of the molecule, allowing the excitation of \revision{photons}. 
We investigate theoretically this system for the experimentally relevant limit of
large damping rate $\kappa$ for the cavity mode and arbitrary coupling 
strength to a single-electronic level. 
We find that for bias voltages close to the first inelastic threshold of photon 
emission, the emitted light 
displays anti-bunching behavior with vanishing second-order photon correlation function. 
At the same time, the current and the intensity of emitted light display
Franck--Condon steps at multiples of the cavity frequency $\omega_c$
with a width controlled by $\kappa$ rather than the temperature $T$.
For large bias voltages, we predict strong photon bunching of the order of  $\kappa/\Gamma$ 
where $\Gamma$ is the electronic tunneling rate.
Our theory thus predicts that strong coupling to 
a single level allows current-driven non-classical light emission.
\end{abstract}

\date{\today}

\maketitle

Electronic transport coupled to the field of an electromagnetic cavity 
can be realized in a wealth of different systems. 
This includes in the microwave range carbon-nanotubes \cite{PhysRevLett.107.256804,BrViDa.16.CavityPhotonsas,CoKoJPCM,PhysRevB.98.155313,cubaynes2019highly}, 
quantum-dots 
\cite{MiCaZa.16.Strongcouplingsingle,doi:10.1063/1.4974536,MiCaZa.16.Strongcouplingsingle,PhysRevX.7.011030,PhysRevLett.113.036801}, 
and
Josephson-junctions 
\cite{wallraff2004strong,PhysRevLett.122.186804,PhysRevX.9.021016}, or in the optical range, 
molecules in plasmonic nanocavities, formed by an STM tip with a substrate
\revision{\cite{berndt_inelastic_1991, QiNaHo.03.VibrationallyResolvedFluorescence, schneider_plasmonic_2012,  ZhZhDo.13.Chemicalmappingof, ReScSp.14.ElectroluminescencePolythiopheneMolecular,
MeGrRo.15.ExcitondynamicsC60,
ZhLuZh.16.Visualizingcoherentintermolecular,ZhYuCh.17.Electricallydrivensingle,
PhysRevLett.119.013901, Doppagne251, ChAfSc.18.BrightElectroluminescenceSingle,  NeEsCa.18.CouplingMolecularEmitters}}
and organic microcavities \cite{OrHuEbnat2015,PhysRevLett.106.196405,PhysRevLett.119.223601}, 
\revision{or with waveguide quantum electrodynamic systems \cite{Chen:16,PhysRevA.96.053805,Zhou:19}.} 
The reduction of the cavity volume $\mathcal{V}$ allows to increase the zero-point 
quantum fluctuations of the electric field $E_{\rm zpm} \sim \mathcal{V}^{-1/2}$.
This motivated optical studies of molecular two-level systems strongly 
coupled to the cavity field by the dipolar interaction $\Lambda_\mathrm{d} \sim p E_{\rm zpm}$,
(with $p$ the molecule dipole moment).
%
One of the goals of this effort is to reach $\Lambda_\mathrm{d}$ larger than 
$\kappa$, which has been and remains challenging, despite recent 
achievements \cite{ChBaJe.16.Singlemoleculestrongcoupling}. 
On the other side, the coupling of a cavity mode 
to the current-induced charge fluctuations of a single-electronic level is given by 
a monopolar coupling constant $\Lambda_\mathrm{m} \sim e L E_{\rm zpm}$ 
\revision{as derived in Ref.~\cite{CoKoDo.15.Electronphotoncoupling}
(see also \cite{supp})},
with $L$ the typical extension of the transport region and $e$ the electronic charge. 
Since typically in a given system 
$eL \gg p$, the monopolar coupling constant is much larger than the dipolar one
\cite{CoKoDo.15.Electronphotoncoupling}.
This probably contributed to the observation 
of values of $\Lambda_\mathrm{m}$
larger than $\kappa$
in microwave cavities coupled to electronic transport 
\cite{MiCaZa.16.Strongcouplingsingle,PhysRevX.7.011030,PhysRevB.98.155313} and even approaching the cavity resonating frequency $\omega_c$ ($\hbar=1$)
\cite{PortierApplPhysLett2013,Cassidy939,PhysRevLett.122.186804}.
Recent results in plasmonic cavities
coupled to electronic transport \revision{\cite{ReScSp.14.ElectroluminescencePolythiopheneMolecular,PhysRevLett.119.013901,
Doppagne251}}
open thus the possibility to explore transport through a single electronic level in these structures.
This is expected to reach much larger coupling constants than those currently
observed for purely dipolar coupling, requiring further theoretical investigations. 
%
\begin{figure}[b]
\includegraphics[width=.9\columnwidth,trim=0 0 0 0,clip]{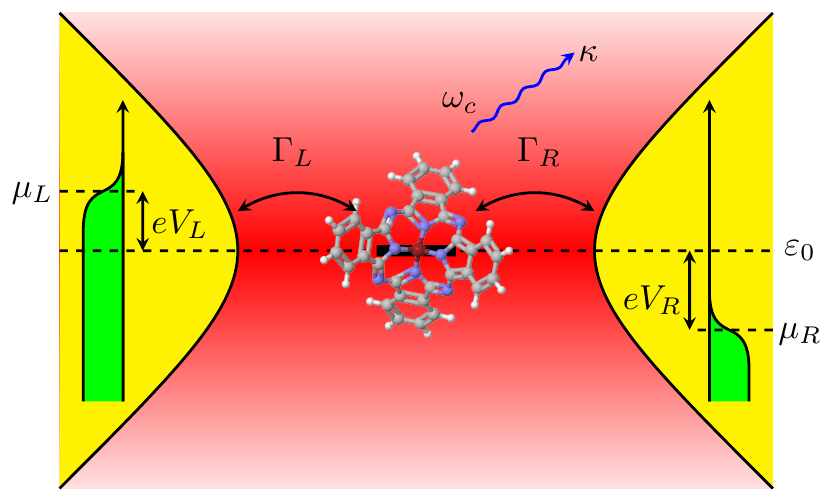}
\caption{
Schematic of two metallic electrodes forming a plasmonic nanocavity characterized by a resonating 
frequency $\omega_c/2\pi$ and damping rate $\kappa$.
%
%
A single electronic level $\varepsilon_0$ of a molecule in the nanogap
couples to the electromagnetic radiation with coupling constant $\Lambda_\mathrm{m}$. 
Electrons can tunnel to and from the dot with \revision{tunneling} rates $\Gamma_\alpha$. 
Voltage drops, $V_\alpha$, with respect to $\varepsilon_0$ are indicated.
}
\label{fig:setup}
\end{figure}
%

The system presents strong analogies with electron-transport coupled to molecular vibrations.
This has been investigated in different regimes, leading to the 
striking prediction of Franck--Condon blockade \cite{PhysRevB.68.205324,Kovo.05.Franck-Condonblockadeand,PhysRevB.74.205438}
and its observation \cite{FCBexp,burzuri2014franck}.
However, there are important differences:
The first is the low quality factor of plasmonic cavities, which is 
typically of the order of 10 \cite{ChBaJe.16.Singlemoleculestrongcoupling}. 
The second, and more interesting, is that the state of the optical 
or microwave cavity can be directly measured by detecting the emitted photons. 
It is thus important to investigate how transport through a molecule is linked to the 
property of the emitted radiation \cite{GaMiNi-prl-2005,KaKrNi-prl-2015,XuHoRaBe-prb-2016}. 

In this paper we consider electronic transport through a single-level quantum dot, 
where the charge on the dot is coupled to the electric field of an electromagnetic cavity. 
We \revision{propose a theoretical model to obtain the}
current through the quantum dot taking into account the cavity dissipation $\kappa$, 
and arbitrary values of the coupling strength 
in the incoherent transport regime $\Gamma\ll k_B T$, with $\Gamma$ the 
electron \revision{tunneling} rate, $T$ the temperature, and $k_B$ the Boltzmann constant. 
Similarly to the Franck--Condon case, we find current steps at the inelastic thresholds for photon emission, but with a width 
controlled by $\kappa$ rather than $T$. 
%
We also derive the photon distribution and the second-order photon correlation function 
$g^{(2)}(t)$, where $t$ is the emission time. 
Its behavior for $t=0$ clearly shows that close to the first threshold 
for photon emission, for $\Gamma/\kappa \ll 1$, and $\Lambda_\mathrm{m}\approx \omega_c$,
photons anti-bunch: 
The junction becomes a single-photon source based on single-electron tunneling. 
\revision{This mechanism is different from the one assumed to be responsible for recently observed anti-bunching of emitted light in STM plasmonic nanojunctions involving multiple electronic levels \cite{ZhYuCh.17.Electricallydrivensingle}.}
For large bias voltages we find instead strong photon bunching 
with $g^{(2)}(0)\approx\kappa/\Gamma \gg 1$.
%
%

{\em Model.}
Figure \ref{fig:setup} shows a schematic of the system at hand.
The Hamiltonian is written
$H=H_{\rm S}+H_{\rm I}+H_{\rm B}$,
where
\begin{equation}
H_{\rm S}
=\tilde\varepsilon_0 d^\dag d^{\phantom\dag}
+
\omega_c a^\dag a + \Lambda_\mathrm{m} d^\dag d (a+a^\dag) ,
\label{eq:hamiltonian}
\end{equation}
with $d^\dag$ the creation operator for the electron on the dot 
single level of energy 
$\tilde\varepsilon_0$,
$a^\dag$ the creation operator for the photon field, and
$\Lambda_\mathrm{m}=\lambda \omega_c$ the coupling constant.
\revision{We follow Ref.\cite{CoKoDo.15.Electronphotoncoupling} for 
the derivation of the interaction term \cite{supp}.
We neglect direct coupling of the cavity field to the electrons 
in the leads, since this effect is analogous to the coupling 
of the cavity field to the photon bath.} 
We treat the electrons in the leads and the propagating 
electromagnetic modes as a bath:
$
H_{\rm B} 
    = \sum_{\alpha k} \varepsilon_{\alpha k} 
    c^\dag_{\alpha k}c^{\phantom\dag}_{\alpha k}
    +
    \sum_q \omega_q b^\dag_q b^{\phantom\dag}_q
$,
where $c^\dag_{\alpha k}$ and $b^\dag_{q}$ are the creation operators for the electrons 
on the leads $\alpha=L,R$ with energy $\varepsilon_{\alpha k} $ and for
the propagating photons of energy $\omega_q$, respectively.
The (linear) coupling to the bath is given by
$
H_{\rm I}=
    \sum_{\alpha k}
    \big[
    t_{\alpha k}c^\dag_{\alpha k}d+t^*_{\alpha k}d^\dag c^{\phantom\dag}_{\alpha k}
    \big]
    +
    \sum_q
    \big[
        l_q a^\dag b^{\phantom\dag}_q+l^*_q b^\dag_q a 
    \big]
$,
with $t_{\alpha k}$ and $l_{q}$ the \revision{tunneling} amplitudes.
We first perform a standard Lang-Firsov unitary transformation on the 
Hamiltonian $\tilde{H}=UHU^\dagger$, with $U=e^{\lambda d^\dag d (a-a^\dag)}$.
%
This removes explicitly the electron-photon coupling term in $H_{\rm S}$, shifts the dot-level energy $\varepsilon_0=\tilde{\varepsilon}_0-\Lambda_\mathrm{m}^2/\revision{\omega_c}$
and modifies the $d$ operator in $H_{\rm I}$ into $D=d e^{\lambda(a-a^\dag)}$.
%
%
\begin{figure}[t]
	\includegraphics[width=.9\columnwidth,trim=0 10 10 10,clip]{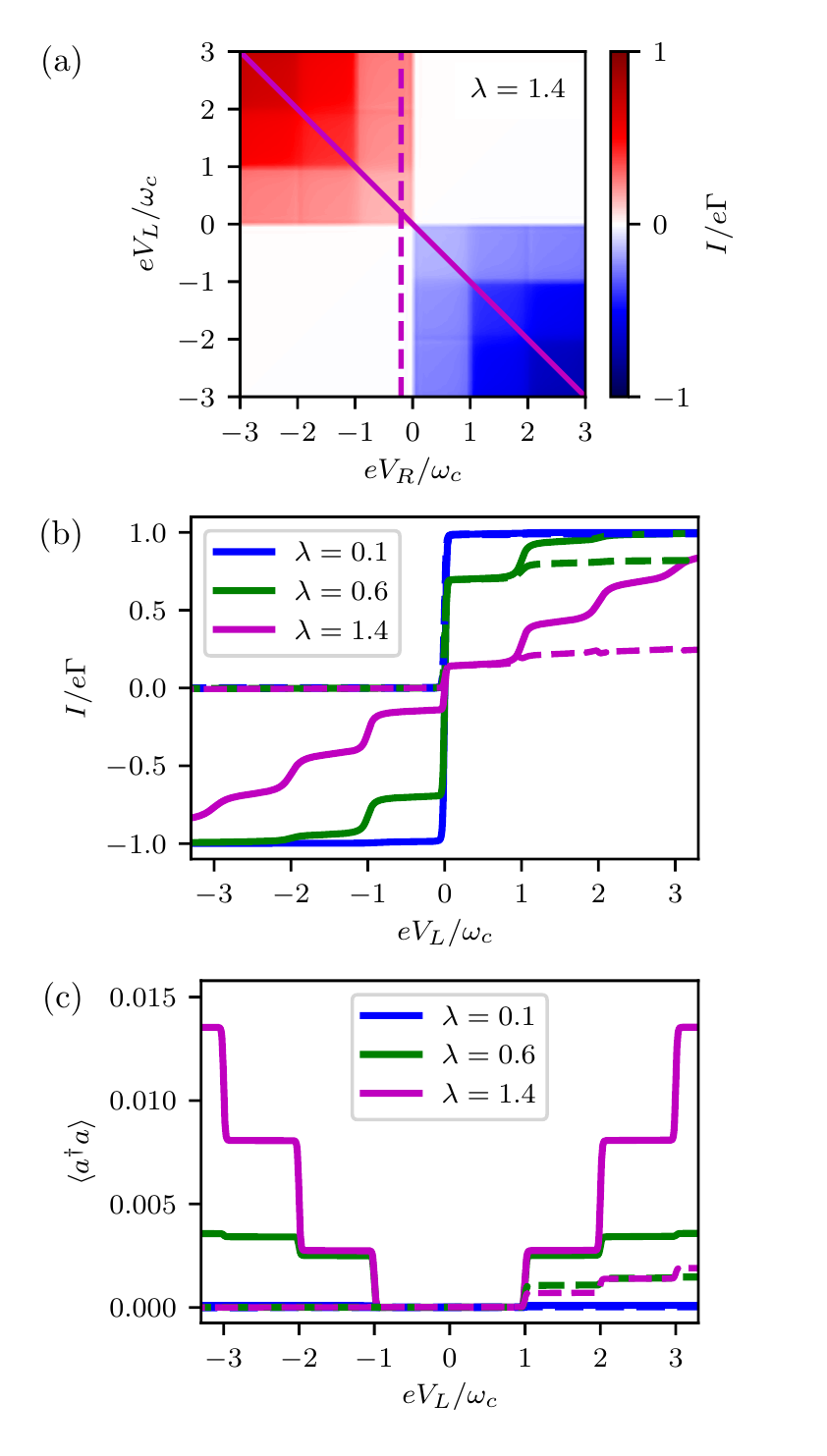} 
	\caption{
	(a) Electronic current $I$ versus the voltage drops $V_L$ and $V_R$ for $\lambda=1.4$.
	(b) Current-voltage characteristics and (c) average photon occupation $\langle a^\dagger a\rangle$ in the cavity corresponding to the voltage profiles indicated in panel a (dashed and full lines)
	for different electron-photon coupling strengths
	$\lambda=0.1$ (blue), $\lambda=0.6$ (green) and $\lambda=1.4$ (magenta).
	The model parameters are
	$\kappa =10k_BT =0.1\omega_c$  and $\Gamma_L=\Gamma_R=\Gamma/2=10^{-3} \omega_c$.
	}
	\label{fig:damped}
\end{figure}

\begin{figure*}[t]
	\includegraphics[width=\textwidth,trim=10 10 0 0,clip]{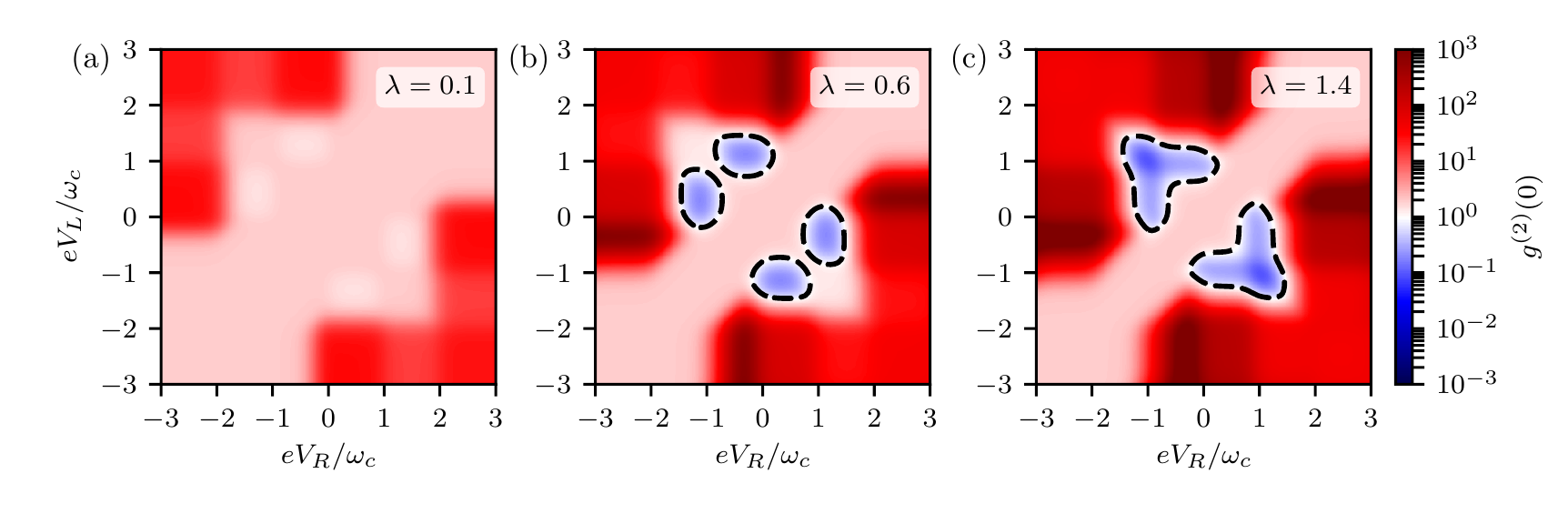}
	\caption{Degree of coherence $g^{(2)}(0)$ for different electron-photon coupling strengths (a) $\lambda=0.1$, (b) $\lambda=0.6$, and (c) $\lambda=1.4$, respectively. 
	The black dashed lines mark the contour $g^{(2)}(0)=1$ delimiting the regions of anti-bunching (blue areas).
	The model parameters are $\kappa=k_BT=0.1\omega_c$  and $\Gamma_L=\Gamma_R=10^{-3}\omega_c$.
	}
	\label{fig:g2}
\end{figure*}


%
{\em Master-equation.} 
Let us define the reduced density matrix $\rho(t)$ for the $d$ and $a$ degrees of freedom 
after tracing out the bath.
\revision{We assume that the molecule is sufficiently isolated
from the substrate, as is reasonable for STM experiments performed on thin insulating films \cite{QiNaHo.03.VibrationallyResolvedFluorescence,
ZhLuZh.16.Visualizingcoherentintermolecular,ZhYuCh.17.Electricallydrivensingle,
PhysRevLett.119.013901,Doppagne251}.
}
We consider then the relevant regime $\Gamma \ll k_B T \ll \omega_c$ 
where
the dynamics of $\rho(t)$ can be described by the Born-Markov master
equation:
\begin{eqnarray}
\dot{\rho}(t) = \mathcal{L}\rho(t)= -i[\tilde{H}_{\rm S}, \rho(t)]+ \left(\mathcal{L}_\mathrm{c}+\mathcal{L}_e\right)\rho(t).
\label{ME}
\end{eqnarray}
The first term contributing to the Liouvilian operator
$
\mathcal{L}
$
gives the coherent evolution of $\rho(t)$.
The second one describes 
the damping of the cavity mode \revision{\cite{Louisell:103059, GaCo.85.Inputoutputdamped}}:
$
\mathcal{L}_\mathrm{c} \rho(t) =
\kappa
\left(
2a \rho(t) a^\dag- a^\dag a \rho(t) -\rho(t) a^\dag a
\right)/2
-
\kappa n_B
\left\lbrack 
a^\dagger,\left\lbrack a, \rho(t) \right\rbrack
\right\rbrack
$,
with 
$
n_B(\omega_c) = \left\lbrace e^{\omega_c/k_B T}-1\right\rbrace^{-1}
$
the Bose distribution of photons in the bath at the cavity frequency.
The last term describes incoherent electron tunneling
$
\mathcal{L}_e\rho(t)=
\left\lbrack \mathcal{D}_-\rho(t)
-\rho(t)\mathcal{D}_+,D^\dag \right\rbrack 
+ \mbox{h.c.}
$  \cite{PhysRevB.97.155414},
where
$
\mathcal{D}_\pm=
\int_{-\infty}^{\infty}d\omega \int_0^{\infty} dt \Gamma_\alpha(\omega) f_\alpha^\pm(\omega)e^{i\omega t}D_{\rm{I}}(-t),
$ and
$
\Gamma_\alpha=2\pi\sum_{k}|t_{\alpha k}|^2\delta(\omega-\varepsilon_{\alpha k})
$ is the tunneling rate from the lead $\alpha$, that in the usual wide-band approximation becomes 
$\omega$-independent. 
Finally, $D_{\rm{I}}(-t)$ is the $D$ operator in the interaction representation 
with respect to $\tilde{H}_\mathrm{S}$.
We introduced the short-hand notation
$f^{+}_{\alpha}\left(\omega\right)=1-f^{-}_{\alpha}\left(\omega\right)=n_F(\omega-\mu_\alpha)$,
with $\mu_\alpha$ the chemical potential of lead $\alpha$ and
$
n_F(\omega) = \left\lbrace e^{\omega/k_B T}+1\right\rbrace^{-1}
$ the Fermi distribution.
The average or the correlation function of any observable $A$, $B$, can then be calculated
in the stationary regime by 
$\left\langle A  \right\rangle = \mathrm{Tr}[A\rho^\mathrm{st}]$ and 
$S_{AB}(t)=\left\langle A(t)B(0)\right\rangle=
\mathrm{Tr}[A e^{\mathcal{L}t} B \rho^\mathrm{st}]$ \cite{cohen1998atom,PhysRevB.86.081305},
with $\rho^\mathrm{st}$ the stationary solution of \Eqref{ME}.

{\em Electronic current.} 
Using the previous results, we derive the expression for the average electronic 
dc-current evaluated at lead $\alpha$
%
\begin{equation}\label{eq:Current}
    I_{\alpha} = \frac{e\Gamma_\alpha}{\pi}\, 
    \mathrm{Re}\!\!\int_{-\infty}^{\infty} \!\!d\omega \left\lbrace f^{+}_\alpha(\omega)S_{DD^\dag}(\omega)
    \!+\! f^{-}_\alpha(\omega)S_{D^{\dag}D}(\omega) \right\rbrace
\end{equation}
where we introduced the Fourier transform of $f^\pm_\alpha$, $S_{DD^\dag}$ and $S_{D^\dag D}$.
Equation (\ref{eq:Current}) enables to calculate the current in presence of strong damping rates $\kappa$, that for plasmonic cavities reaches low quality factors $\omega_c/\kappa\approx 10$ \cite{ChBaJe.16.Singlemoleculestrongcoupling}.
%
It allows to include the damping of the electromagnetic field during the \revision{tunneling} process.
This expression and $\rho^{\rm st}$ can be evaluated numerically by projecting 
on the charge and harmonic oscillator basis.

Figure \ref{fig:damped}(a) reports the electronic current $I$ 
for strong coupling, $\lambda=1.4$,
as a function of the relative voltage drops $eV_\alpha=\mu_\alpha-\varepsilon_0$ between the chemical potential of lead $\alpha$ and the dot energy level \cite{current-conservation}.
Specific current-voltage characteristics, corresponding to symmetric ($V_L=-V_R$) and asymmetric ($eV_R=-0.2\omega_c$) voltage drops, are shown respectively as full 
and dashed lines in \Figref{fig:damped}(b), for weak ($\lambda=0.1$), 
moderate ($\lambda=0.6$), and strong ($\lambda=1.4$) coupling strengths.
These exhibit similar features of the Franck--Condon blockade regime \cite{Kovo.05.Franck-Condonblockadeand,PhysRevB.74.205438,PhysRevB.68.205324}, 
with inelastic steps observed each time the voltage drop $eV_L=n\omega_c$ matches a multiple of the cavity-photon frequency.
This is the threshold for one-electron tunneling while emitting $n$ photons in the cavity.
The step heights are given by the Poisson distribution 
$\mathcal{P}_n(\lambda)=e^{-\lambda^2}\lambda^{2n}/n!$ 
\revision{
and the width of the inelastic steps by the cavity-losses $\kappa$, 
which exceed the temperature broadening. 
This is analogous to the broadening of phonon sidebands by frictional damping 
\cite{PhysRevB.68.205324}, but our treatment is not bound to thermal equilibrium.}

{\em Emitted light.} 
We consider now the emitted light power $\sim \kappa \omega_c \langle a^\dag a\rangle$, 
by plotting the average population of the cavity mode in \Figref{fig:damped}(c) 
as a function of $V_L$. 
We find that the photon population also increases with bias voltage in a step-like manner \cite{broadening}, 
correlated to the evolution of the electronic current \cite{PhysRevB.97.155414},
thus confirming that single-electron tunneling is at the origin 
of light-emission inside the cavity. 
From \Eqref{ME}, performing a secular approximation, 
we derive a rate equation for the photon population $P_n(t)$ \cite{supp}.
A relevant experimental regime in plasmonic cavities is $\kappa \gg \Gamma$ and 
$\omega_c \gg k_BT$.
%
%
Since the time between two \revision{tunneling} events is much longer than the damping time 
of the cavity, 
typically the circulating photon leaks out before a new photon is emitted in the cavity.
In this limit $P_0 \approx 1$, and we find for the other populations
$P_n=\Gamma_{0n}/n\kappa \ll 1$,
with 
$\Gamma_{0n}=\sum_\alpha \Gamma_\alpha \mathcal{P}_n(\lambda)\, n_F(n\omega_c-\mu_\alpha)$
the cavity 0 to $n$-photons transition rate induced by a single \revision{tunneling} event. 
The expression for $\langle a^\dag a\rangle=\sum_n \Gamma_{0n}/\kappa$ describes then accurately the emitted power. 
%

\begin{figure*}[t!]
    \includegraphics[width=\textwidth,trim=6 10 0 0,clip]{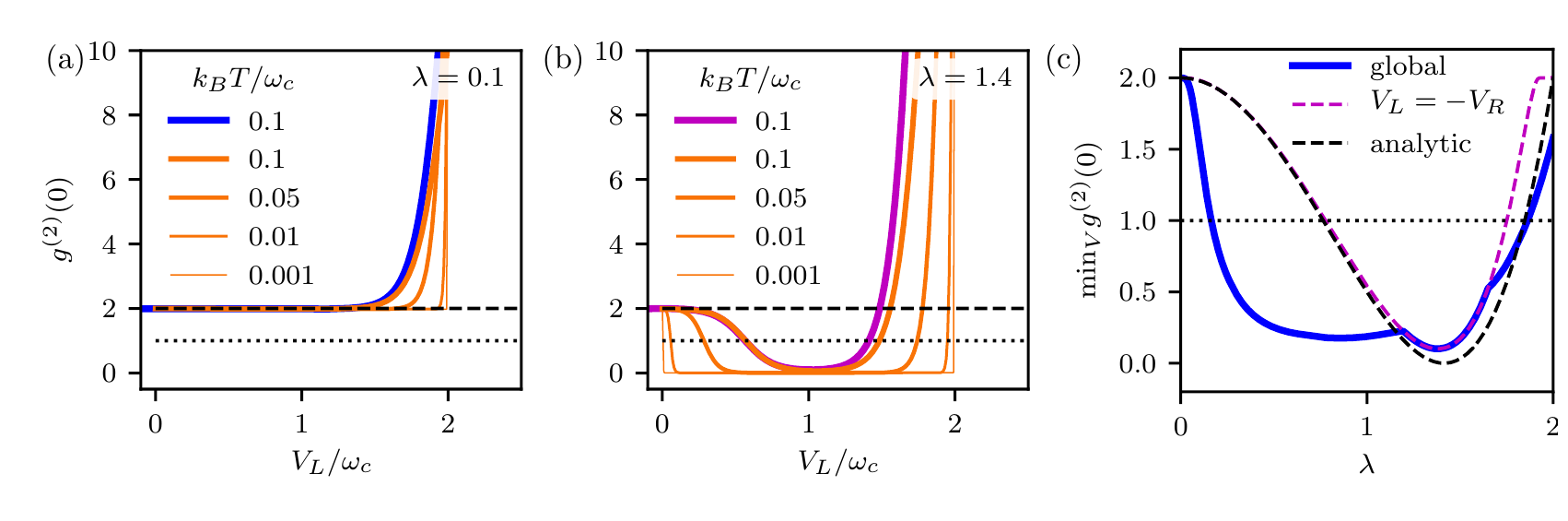}
	\caption{
	(a), (b) Role of temperature $T$ on the degree of coherence $g^{(2)}(0)$ for $V_L=-V_R$ with (a) $\lambda=0.1$ and (b) $\lambda=1.4$.
	%
	Blue and magenta thick lines are \revision{slices from \Figref{fig:g2}(a,c)}, while the orange ones are computed from \revision{rate equations \cite{supp}}.
	%
	%
	Dashed (dotted) horizontal line 
	indicates $g^{(2)}(0)=2$ $(1)$ for thermal (uncorrelated) photon emission.
	%
	%
	(c) Minimum value of $g^{(2)}(0)$ (thick blue line) on the plane $(V_L,V_R)$, for fixed $\lambda$ and $k_B T=0.1 \omega_c$.
	Dashed magenta line shows that minimum constrained to $V_L=-V_R$.
	Dashed black line gives the $T=0$ analytical prediction $(\lambda^2-2)^2/2$.
	Other parameters are those of \Figref{fig:g2}.
	%
	%
	}
	\label{fig:g2-lowT}
\end{figure*}

%
{\em Correlation function $g^{(2)}$.} 
In order to characterize the statistics of the emitted light, 
we compute the second-order correlation function
$g^{(2)}(t)=\langle a^\dag a^\dag(t)a(t) a\rangle/\langle a^\dag a\rangle^2$
\cite{nla.cat-vn4402306,Louisell:103059,RiCa1988,PhysRev.131.2766}. 
Let us begin with the $t=0$ case, $g^{(2)}(0)=(\sum_n n(n-1) P_n)/(\sum_n n P_n)^2$. 
%
One can readily verify that in thermal equilibrium $g^{(2)}(0)=2$.
On \Figref{fig:g2} we show    
$g^{(2)}(0)$ as a function of $V_R$ and $V_L$, for three different values of $\lambda$. 
%
As expected, for $e|V_L-V_R| \ll k_B T$, one always finds the value of 2, corresponding to 
thermal equilibrium (pink regions on the diagonal $V_L=V_R$). 
Out of equilibrium we find either bunching $g^{(2)}(0) > 1$ 
or anti-bunching $g^{(2)}(0)<1$. 
The anti-bunching appears for sufficiently strong coupling and 
is indicated by the blue regions with dashed border 
(where $g^{(2)}(0)=1$).
%
\revision{
Treating both the electron tunneling and the thermal excitations
as weak perturbation to the distribution $P_n=\delta_{n,0}$ we 
obtain an analytical expression for $g^{(2)}(0)$ (see \cite{supp})
for voltages $eV_L=-eV_R \leq 2 \omega_c$.}

The prediction of this expression agrees very well 
with the numerical calculations 
for $k_B T \geq 0.1 \omega_c$, (cf.~thick lines in 
\Figref{fig:g2-lowT}(a) and (b)).
At lower temperature we thus show only the result obtained 
with the analytical expression, 
that does not \revision{suffer from} numerical instability.
In the strong coupling regime,  
we predict a smooth crossover 
from the equilibrium value $g^{(2)}(0)=2$ at low voltage, to an anti-bunching $g^{(2)}(0)<1$ regime that appears close to the first inelastic threshold ($|eV_L| \approx \omega_c$ or $|eV_R|\approx \omega_c$) \revision{\cite{supp,PhysRevA.41.475}}.
As temperature decreases the region of anti-bunching expands, 
eventually including almost 
the full bias range 
$\left\lbrack 0,  2\omega_c\right\rbrack$,
cf.~\Figref{fig:g2-lowT}(b).
Figure \ref{fig:g2-lowT}(c) shows the minimum value taken
by $g^{(2)}(0)$ when minimized on the plane $(V_L,V_R)$
for a given value of the coupling $\lambda$.
One finds that anti-bunching can be observed for $\lambda>0.17$ for 
an asymmetric bias configuration, thus being 
at reach of present experiments.
The case of symmetric bias is also shown, with 
anti-bunching beginning at $\lambda>0.76$.
Let us discuss now the anti-bunching mechanism for symmetric bias. 
When $P_{n+1}\ll P_n$, at lowest order 
$g^{(2)}(0)=2P_2/P_1^2$.
This expression gives $2$ for 
thermal distribution $P_n\sim e^{-n\omega_c/k_BT}$.
Anti-bunching is thus achieved for $2 P_2 \ll P_1^2$. 
At very low temperature 
the only way to populate the state $2$ is either 
a 2-photons transition (for $eV_L>2\omega_c$), or an electron-tunneling 
assisted transition from the state 1 to the state 2, 
controlled by 
$\Gamma_{12}=
e^{-\lambda^2} \lambda^{2}
\left( 2 - \lambda^2
\right)^2
\sum_\alpha \Gamma_\alpha n_F(\omega_c-\mu_\alpha)/2 
$
(for $eV_L > \omega_c$).
%
As a confirmation one finds that 
$P_1 \approx P_0 \Gamma_{01}/\kappa$ and 
$P_2 \approx P_1 \Gamma_{12}/2\kappa$, 
testifying that the result is just due to a balance
between the light leaked out of the cavity
and 
the photons emitted in the cavity by the \revision{tunneling} electrons. 
From the explicit expression of $\Gamma_{12}$ 
at low temperature ($k_BT \ll \omega_c$)
one then finds that the minimum value of $g^{(2)}(0)$ is approximated by $\left( \lambda^2 - 2 \right)^2/2$ 
[dashed black line in \Figref{fig:g2-lowT}(c)], tracing the $\lambda$-dependence of $\Gamma_{12}$. 
Its vanishing for $\lambda=\sqrt{2}$ is thus at the origin of the anti-bunching. %
A similar effect has also been recently reported in dc-biased Josephson 
junction coupled to microwave resonators \cite{PhysRevLett.122.186804,PhysRevX.9.021016}.
For larger voltages $eV_L \ge 2\omega_c$, we obtain analytically
\cite{supp}
$g^{(2)}(0) \approx \kappa / 2\Gamma$, as predicted by the 
numerical simulations giving a 
smooth evolution to strong bunching. 
This result agrees with the infinite bias voltage limit result 
recently reported in Ref.~\cite{BergSamuelsson2019}.
Finally the time dependence of $g^{(2)}(t)$ as obtained by the numerical 
calculations shows a smooth crossover on a time scale $1/\kappa$ 
from $g^{(2)}(0)=2$ to $g^{(2)}(\infty)=1$ (uncorrelated photons) \cite{RiCa1988}.
%
Only for $\lambda\approx 1$ we observe weak oscillations \cite{supp}.

{\em Conclusions.} 
Charge fluctuations  
induced by electronic transport in a molecular {\em single}-electronic level
are expected to couple strongly to  
the plasmonic mode formed by an STM tip and the substrate. 
We derived an expression for the current taking into account the strong 
damping of these cavities and obtained the current, emitted light intensity, and the correlation function $g^{(2)}(t)$.
We showed that when the coupling strength is of the order $\lambda\sim 1$,
Franck--Condon steps appear in 
both the current and the light intensity. 
Non-classical light can be emitted for a coupling strength in the range 
$0.17 < \lambda < 1.8$ for bias voltage near the one photon emission threshold.
This prediction can be relevant for a series of experiments on STM cavities 
\revision{\cite{ReScSp.14.ElectroluminescencePolythiopheneMolecular, ZhZhDo.13.Chemicalmappingof, 
MeGrRo.15.ExcitondynamicsC60, ZhLuZh.16.Visualizingcoherentintermolecular, Doppagne251, ChAfSc.18.BrightElectroluminescenceSingle, ZhYuCh.17.Electricallydrivensingle, PhysRevLett.119.013901,
NeEsCa.18.CouplingMolecularEmitters}}.
\revision{The importance of single-level descriptions has further been reinforced experimentally after our initial submission \cite{LeGuOt.19.SinglePhotonEmission}.}
%
\revision{
The fast evolution of the cavity ($\kappa>k_B T$) might question the validity of the Markov approximation. 
We estimate that the non-Markovian contributions are negligible 
as far as $k_B T \ll \omega_c$, but investigating low-frequency 
response of plasmonic cavities could unravel such effects.}
Another interesting direction is the study of current response in presence 
of light irradiation of the plasmonic junction. 
\revision{Finally, we note that a local quantum emitter can be used to sense the environment of the 
molecule with the minimal quantum of energy.} 
\begin{acknowledgments} 
{\em Acknowledgments.}We thank Tom{\'a}\v{s} Neuman, Javier Aizpurua, and Guillaume Schull for stimulating discussions.
This work was supported by IDEX Bordeaux (No.~ANR-10-IDEX-03-02) and Euskampus Transnational Common Laboratory \textit{QuantumChemPhys}.
T.F. acknowledges grant~FIS2017-83780-P from the Spanish Ministerio de Econom{\'i}a y Competitividad.
R.A. acknowledges financial support by the Agence Nationale de la Recherche project CERCa, ANR-18-CE30-0006.
\end{acknowledgments}
\phantom{\cite{PhysRevA.41.475}}

\bibliographystyle{apsrev4-1}
\bibliography{references.bib}

\begin{thebibliography}{57}
\bibitem [{cur()}]{current-conservation}%
  \BibitemOpen
  \href@noop {} {}\bibinfo {howpublished} {We plot the symmetrized current
  $I=(I_L+I_R)/2$, since for asymmetric bias voltages one observes in some
  cases weak non-conserving contributions.}\BibitemShut {Stop}%
\bibitem [{bro()}]{broadening}%
  \BibitemOpen
  \href@noop {} {}\bibinfo {howpublished} {At this level of approximation, the
  steps of the cavity-mode occupation appear to be broadened on the scale given
  by temperature instead of $\kappa$.}\BibitemShut {Stop}%
\end{thebibliography}

\clearpage

\end{document}


\title{Supplementary Material \--- Single-photon emission mediated by single-electron tunneling in plasmonic nanojunction}
\author{Q.~Schaeverbeke}
\affiliation{Univ.~Bordeaux, CNRS, LOMA, UMR 5798, F-33405 Talence, France}
\affiliation{Donostia International Physics Center (DIPC), E-20018, Donostia-San Sebasti\'an, Spain}
\author{R.~Avriller}
\affiliation{Univ.~Bordeaux, CNRS, LOMA, UMR 5798, F-33405 Talence, France}
\author{T.~Frederiksen}
\affiliation{Donostia International Physics Center (DIPC), E-20018, Donostia-San Sebasti\'an, Spain}
\affiliation{Ikerbasque, Basque Foundation for Science, E-48013, Bilbao, Spain}
\author{F.~Pistolesi}
\affiliation{Univ.~Bordeaux, CNRS, LOMA, UMR 5798, F-33405 Talence, France}

\begin{abstract}
We provide supplementary information about the microscopic Hamiltonian describing the coupling between the plasmon mode of a STM cavity and the electronic single-level of a molecule embedded between the STM tip and the substrate. 
The dependence of the tunneling current across the junction with the plasmon-molecule coupling strength is analyzed, and shows analogous features to the Franck-Condon blockade regime. 
The second-order correlation function of the cavity plasmon field is investigated further both analytically and numerically, in the full bias-voltage range and in a regime where dissipation $\kappa$ of the cavity is larger than the tunneling rate $\Gamma$.
\end{abstract}

\date{\today}

\maketitle

\section{Coupling of the plasmon mode to the electronic nanojunction}

\subsection{Microscopic Hamiltonian}

In this section, we provide more details about the microscopic
Hamiltonian describing the coupling between a STM local plasmon mode, and a single  molecule embedded between the STM tip and the substrate.
The form of the interaction Hamiltonian in Coulomb gauge is derived in Eq.~(28) of Ref.~\cite{CoKoDo.15.Electronphotoncoupling}, for the case of a microwave cavity mode coupled to a quantum dot connected to a single electronic reservoir. 
Similarly to this case, the interaction Hamiltonian between the electronic STM nanocircuit (including two electronic leads and the electronic single-level of the molecule) and the cavity plasmon mode is given by \cite{CoKoDo.15.Electronphotoncoupling}
\begin{eqnarray}
\hat{H}_{\rm int}&=&\hat{h}_{\rm int}\left( a + a^\dagger\right) \, ,
\label{InteractionH1} \\
\hat{h}_{\rm int} &=& \Lambda_\mathrm{m} d^\dagger d 
+ \sum_{\alpha k} \left\lbrace  
\gamma_{\alpha k}d^\dagger c_{\alpha k} + \gamma_{\alpha k}^* c_{\alpha k}^\dagger d
+
\Lambda_{\alpha k} c_{\alpha k}^\dagger c_{\alpha k}
\right\rbrace \, .
\nonumber \\
\label{InteractionH2}
\end{eqnarray}
The first term in \Eqref{InteractionH2} stands for the direct (monopolar) coupling between the plasmon mode and charge fluctuations of the dot-level.
The corresponding coupling strength is given by the matrix element 
\begin{equation}
\Lambda_\mathrm{m} = -e \int d^3{r} |\phi_d\left(\vec{r}\right)|^2 V_\perp\left(\vec{r}\right) \, ,
\label{InteractionH3} 
\end{equation}
with $\phi_d\left(\vec{r}\right)$ the dot electronic wave function, 
and $V_\perp\left(\vec{r}\right)=-i\omega_c \int_{\mathcal{C}\left( \vec{r}_0,\vec{r} \right)} \vec{A}(\vec{l})\cdot d\vec{l}$ the photonic pseudo-potential \cite{CoKoDo.15.Electronphotoncoupling} obtained as the work of the cavity electrical field
$-i\omega_c \vec{A}(\vec{l})$ ($\vec{A}$ is the vector potential) on a path $\mathcal{C}\left( \vec{r}_0,\vec{r} \right)$ connecting 
any reference point $\vec{r}_0$ to the point $\vec{r}$.
The second term in \Eqref{InteractionH2} describes modulation of the electronic 
tunneling from the leads to the dot induced by the photonic potential.
The corresponding matrix element for this coupling is 
\begin{equation}
\gamma_{\alpha k}=-e \int d^3{r} \phi^*_d\left(\vec{r}\right)\phi_{\alpha k}\left(\vec{r}\right) V_\perp\left(\vec{r}\right) \, ,
\label{InteractionH4} 
\end{equation}
with $\phi_{\alpha k}\left(\vec{r}\right)$ the Bloch function of the metallic lead 
$\alpha=L,R$ with quasi-momentum $k$.
Finally, the last term of \Eqref{InteractionH2} stands for the direct coupling between the electrons in the leads and the cavity plasmon mode
\begin{equation}
\Lambda_{\alpha k}=-e \int d^3{r} |\phi_{\alpha k}|^2\left(\vec{r}\right) V_\perp\left(\vec{r}\right) \, .
\label{InteractionH5} 
\end{equation}

\subsection{Orders of magnitude}

The direct (monopolar) term proportional to $\Lambda_\mathrm{m}$ (see \Eqref{InteractionH2}) contributes to shifting the dot-level energy by an amount $\delta\varepsilon_0=\varepsilon_0-\tilde{\varepsilon}_0=-\Lambda_\mathrm{m}^2/\hbar\omega_c$.
A simple estimation of the coupling strength $\Lambda_\mathrm{m}$ is obtained when neglecting the variation of the photonic pseudo-potential on the scale of the typical extension of the molecular orbitals 
$\Lambda_\mathrm{m} \sim  -e V_\perp\left(\vec{r}_{\rm m}\right) \approx e L E_{\rm zpm}$, with $\vec{r}_{\rm m}$ the location of the molecule, 
$L$ the typical extension of the transport region, and 
$E_{\rm zpm}$ zero-point vacuum fluctuations of the cavity electric field. %

The second term proportional to $\gamma_{\alpha k}$ (see \Eqref{InteractionH2}) provides a contribution to the damping 
of the cavity $\kappa^{(2)}$, obtained at the Fermi golden rule level
by $\kappa^{(2)}=2\pi/\hbar \sum_{\alpha k} |\gamma_{\alpha k}|^2 \delta \left( \tilde{\varepsilon}_0 + \omega_c - \varepsilon_{\alpha k} \right)$.
Using \Eqref{InteractionH4}, we estimate $\gamma_{\alpha k} \approx t_{\alpha k}\Lambda_\mathrm{m}/E_{\rm F}$, with $E_{\rm F}$ the typical Fermi energy of the metallic leads.
This rate is thus of order $\kappa^{(2)} \approx \left( \Lambda_\mathrm{m}/E_{\rm F} \right)^2 \Gamma$.
For the highest coupling strength investigated in the paper, $-\delta \varepsilon_0 \approx \Lambda_\mathrm{m}\approx \hbar \omega_c \approx 2 \mbox{ eV}$ while $E_{\rm F} \approx 5-10 \mbox{ eV}$ for metals, so that 
$\kappa^{(2)} \approx 0.04-0.16 \Gamma \le \Gamma \ll \kappa, -\delta \varepsilon_0$.
%
%
%
The second term in \Eqref{InteractionH2} is thus negligible compared to the monopolar term and to the losses $\kappa$ of the cavity.
The third and last term in \Eqref{InteractionH2} stands for processes (allowed by charge conservation) by which electrons in the leads directly decay to the cavity plasmon mode.
We have not included explicitly this term in the Hamiltonian written in the paper, since it is equivalent to an additional dissipation channel for the plasmon mode, already taken into account by our phenomenological damping term $\propto \kappa$.
%
%

%
\textit{We thus have shown that the monopolar term provides an important (even dominant) contribution to the interaction Hamiltonian in \Eqref{InteractionH2} describing the coupling of a single-electronic level to the cavity plasmon mode.}
%

\section{Electronic tunneling current}
%
\subsection{Photonic analogue to the Franck-Condon blockade}

\begin{figure}[htb]
	\includegraphics[width=\columnwidth,trim=5 10 10 10,clip]{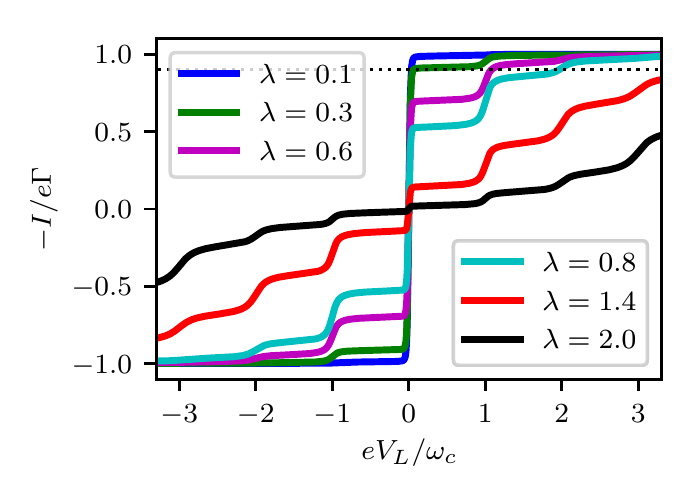}
	\caption{
	Current as a function of the left voltage drop $eV_L$ for a fully symmetrical junction. $V_L=-V_R$, $\Gamma_L=\Gamma_R$, $T=0.01\omega_c$, $\kappa=0.1\omega_c$.
	The black dotted line indicates the value $0.9$ corresponding to $90\%$ of the maximum value of the current.
	}
	\label{fig:IV-plot-IR}
\end{figure}
We present in \Figref{fig:IV-plot-IR} the electronic current $I$ as a function of the voltage drop $eV_L$ in a symmetrical junction ($V_L=-V_R$) for various values of the coupling strength $\lambda$. 
We show that $I(V_L)$ presents steps each times the voltage drop reaches a n-photon emission threshold $V_L \approx n\omega_c$.
The height of those steps increases with $\lambda$ and after reaching a maximum, decreases again.
Furthermore, \Figref{fig:IV-plot-IR} shows that as $\lambda$ increases, the current gets blocked at low voltages, a phenomenon sharing strong analogies to the Franck-Condon blockade regime observed in molecular junctions with phonons \cite{Kovo.05.Franck-Condonblockadeand, burzuri2014franck}.
\begin{figure}[htb]
	\includegraphics[width=\columnwidth,trim=0 10 10 10,clip]{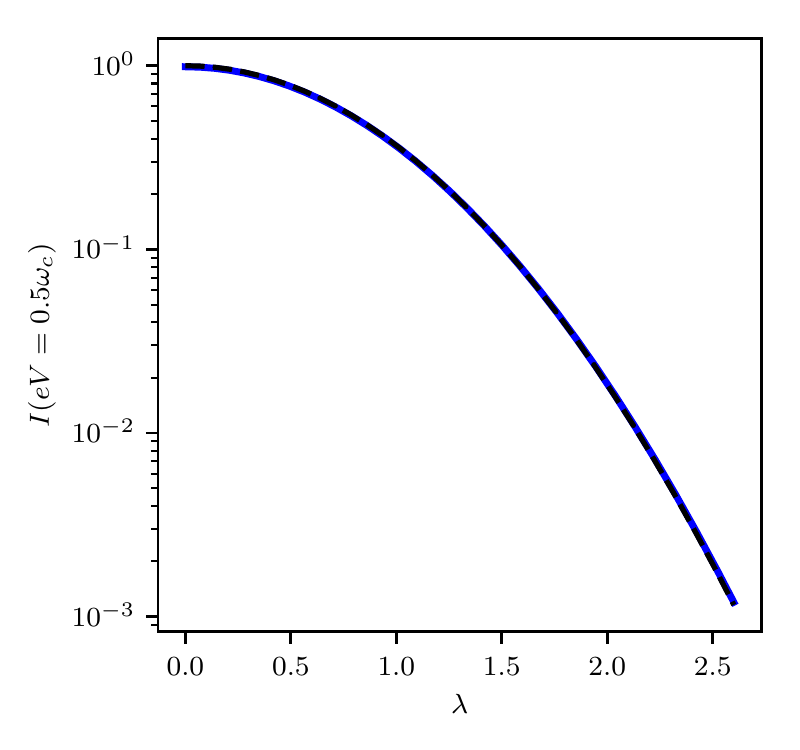}
	\caption{
	Plain blue curve: Value of the current $I$ at $eV_L=0.5\omega_c$ as a function of the coupling strength $\lambda$ in the case of a fully symmetric junction ($V_L=-V_R=V$, $\Gamma_L=\Gamma_R=0.001\omega_c$ and $k_BT=0.01\omega_c$).
	In dashed black is represented the expected dependence $\ln \left( I/\Gamma \right)\approx -\lambda^2$ provided by \Eqref{FC1}.
	}
	\label{fig:I_05}
\end{figure}
We present in \Figref{fig:I_05} the evolution of the electronic current $I$ at low-voltage $eV_L=0.5\omega_c$ with coupling strength $\lambda$. 
Using Franck-Condon blockade theory at low temperature ($T \ll \omega_c$) and in the equilibrated regime for the photon population \cite{Kovo.05.Franck-Condonblockadeand}, we expect the current to scale as 
\begin{equation}
I \approx \Gamma e^{-\lambda^2} \sum_{n=0}^{+\infty} \frac{\lambda^{2n}}{n!}n_F\left( n\omega_c - V_\alpha \right).
\label{FC1}
\end{equation}
Clearly, while \Eqref{FC1} properly predicts the appearance of inelastic thresholds in the current $I(V_L)$, when $V_\alpha \approx n\omega_c$.
It also predicts a width of the steps that is proportional to temperature $T$, in contrast to \Figref{fig:IV-plot-IR} which exhibits steps with a width given by the dissipation rate of the cavity $\kappa$.
This is clearly due to the fact that \Eqref{FC1} does not properly include a description of cavity losses. 
However, some qualitative features present in \Figref{fig:IV-plot-IR} are preserved by \Eqref{FC1}.
For instance, for voltages below the first inelastic threshold $V_L < \omega_c$, \Eqref{FC1} predicts an exponential suppression of the low-bias current: 
$\ln \left( I/\Gamma \right)\approx -\lambda^2$, which is indeed obtained in 
\Figref{fig:I_05}.
Also, \Eqref{FC1} predicts that the typical size of the first inelastic step 
$\Delta I = I\left(V_L = \omega_c + \eta_V\right)-I\left(V_L = \omega_c - \eta_V\right)$, with $\eta_V$ a voltage of order $\kappa$, scales as 
$\Delta I \approx \Gamma \lambda^2 e^{-\lambda^2}$, thus reaching a maximum value 
for $\lambda \approx 1$, which is consistent with the curves of \Figref{fig:IV-plot-IR}.
To show further the impact of $\lambda$ on the current suppression, we plot in \Figref{fig:Vlim} the voltage drop $V_{90}$ needed to recover $90\%$ of the maximum current as a function of $\lambda$.
\Figref{fig:Vlim} shows that $V_{90}$ increases with $\lambda$ in a parabolic manner: 
the higher $\lambda$, the stronger the current suppression at low-voltage, and the higher the needed voltage to reach saturation.
It is interesting to notice that the classical analogue to Franck-Condon blockade theory \cite{PhysRevB.76.165317} predicts that a critical voltage $2\omega_c \lambda^2$ is needed to unblock the junction, which scales accordingly to $V_{90}$ in \Figref{fig:Vlim}.
%

\subsection{Expression of the dc-tunneling current}

The current operator across lead $\alpha=L,R$ is obtained from the continuity equation for tunneling charges as
\begin{equation}
    I_\alpha=ie\sum_{k}\left(t_{\alpha k}c^\dag_{\alpha k}D-
    t^*_{\alpha k}D^\dagger c_{\alpha k}\right),
    \label{Current1}
\end{equation}
with $e$ the elementary charge. 
The average current at time $t$ is computed using the density matrix 
$\underline{\rho}(t)$ of the full system, including the molecular junction with its electronic reservoirs, and the cavity mode with the external photon bath
\begin{equation}
    \langle I_\alpha(t)\rangle=\mathrm{Tr}\left(I_\alpha \underline{\rho}(t)\right)\, .
     \label{Current2}
\end{equation}
The full density matrix in \Eqref{Current2} evolves in time with respect to the 
full Hamiltonian $\tilde{H}$ (after the Lang-Firsov transformation has been performed) as
\begin{eqnarray}
    \dot{\underline{\rho}}(t)&=&-i\left[\tilde{H},\underline{\rho}(t)\right]
    \label{Current3} \, , \\
    \underline{\rho}(0) &=& \rho_L\otimes \rho_R \otimes \rho_b \otimes \rho(0)\, .
    \label{Current4}
\end{eqnarray}
The initial condition for the density matrix in \Eqref{Current4} is chosen to be 
the product of density matrices of the electronic ($\rho_{\alpha=L,R}$) and bosonic ($\rho_b$) environments each supposed to be at equilibrium, times the initial reduced density matrix $\rho(0)$ of system (dot plus cavity-mode).
We further define the "bare" Hamiltonian $H_0=\tilde{H}_{\rm{S}}+\tilde{H}_{\rm{B}}=\tilde{H}-\tilde{H}_{\rm{I}}$ containing the system and baths Hamiltonians.
The "perturbation Hamiltonian" $V=\tilde{H}_{\rm{I}}$ contains the tunneling Hamiltonian (dressed by photons) and the dissipative coupling between the cavity-mode and the photon bath. 
Integrating \Eqref{Current3} with initial condition \Eqref{Current4}, we get
an exact expression for the average current 
\begin{equation}\label{Current5}
    \langle I_\alpha(t)\rangle=\mathrm{Tr}\left(I_\alpha(t)\underline{\rho}(0)\right) -i\int_{t_0}^t dt'\mathrm{Tr} \left( I_\alpha(t)
    \left[V(t'),\underline{\rho}(t')\right]\right) \, ,
\end{equation}
with all operators written in interaction picture with respect to $H_0$.
We further perform the Born approximation $\underline{\rho}(t') \approx \rho_L\otimes \rho_R \otimes \rho_b \otimes \rho(t')$ and truncate consistently \Eqref{Current5} at lowest non-vanishing order in the tunneling rates $\Gamma_\alpha$.
We note that the first term in \Eqref{Current5} vanishes due to our choice of initial conditions.
We further expand the second term after injecting the expression of the interaction Hamiltonian $V$ written in the paper. 
After coming back to Schr\"odinger representation, 
a subsequent Markov approximation enables to neglect the time dependence of $\rho(t')\approx \rho(t)$ in the time integral, and to take the limit $t_0 \rightarrow -\infty$.
We obtain the final expression for the average dc-current
$\langle I_\alpha\rangle=\lim_{t\rightarrow + \infty}\langle I_\alpha(t)\rangle$
\begin{equation}
    \begin{aligned}
    \langle I_{\alpha}\rangle = \frac{e\Gamma}{\pi}\rm{Re} &\int_{-\infty}^{+\infty}d\omega \int_{0}^{+\infty}d\tau
    \left\lbrace f^+_\alpha(\omega)
    \left\langle D^{\phantom\dag}(\tau)D^{\dag}\right\rangle_0 e^{i\omega \tau}
     \right.
    \\
    &\left.-f_\alpha^-(\omega)
    \left\langle D^{\dag}(\tau)D^{\phantom\dag}\right\rangle_0
    e^{-i\omega\tau}\right\rbrace \, ,
    \label{Current6}
    \end{aligned}
\end{equation}
with $\left\langle A(\tau)B\right\rangle_0 = \mathrm{Tr}\left\lbrack e^{iH_0 \tau} A e^{-iH_0 \tau} B \underline{\rho}^{st}\right\rbrack$ the average value of the system operators $A(\tau)$ and $B(0)$ written in interaction representation, using as initial condition the stationary density matrix $\underline{\rho}(0)\equiv \underline{\rho}^{st}$.
%
%

%
In order to incorporate the contribution of cavity dissipation to the time evolution of the $D$ and $D^\dagger$ operators in \Eqref{Current6}, we further replace each term of the form $\left\langle A(\tau)B\right\rangle_0$ by the correlation function $S_{AB}(\tau) = \mathrm{Tr} \left\lbrack e^{i \tilde{H} \tau} A e^{-i \tilde{H} \tau} B \underline{\rho}^{st}\right\rbrack \equiv \mathrm{Tr}_{\rm{S}} \left\lbrack  A e^{\mathcal{L}\tau} B \rho^{st}\right\rbrack$.
\textit{This ansatz is strictly speaking beyond Born approximation, being all orders in the damping rate $\kappa$.}
Equation (3) in the main text of our paper follows after Fourier transformation of the correlation functions
$S_{D D^\dagger}(\tau)$ and $S_{D^\dagger D}(\tau)$.
%

\section{Analytical calculation of $g^{(2)}(0)$}
\begin{figure}[htb]
	\includegraphics[width=\columnwidth,trim=0 10 10 10,clip]{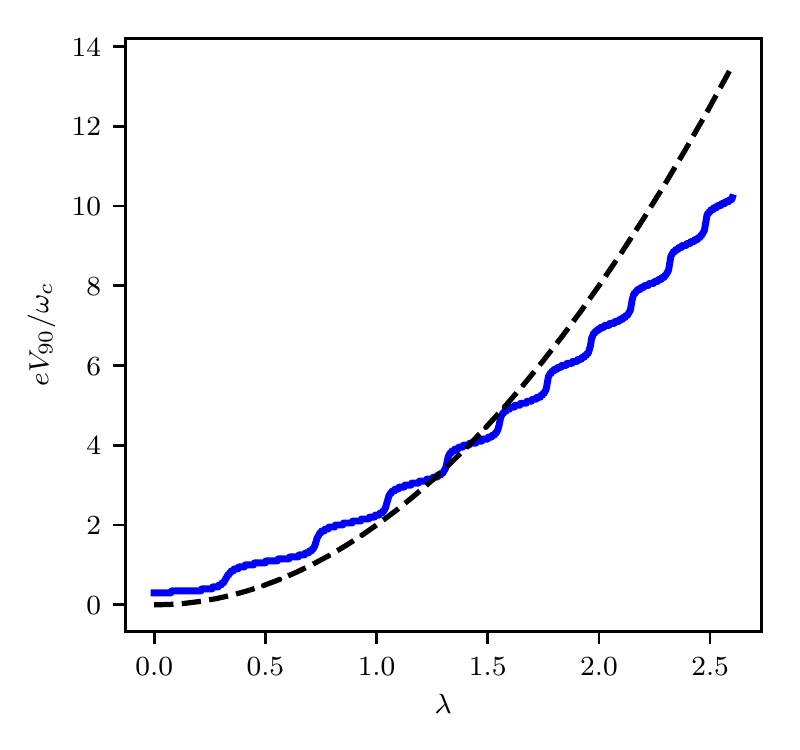}
	\caption{
	Value of $V$ for which $90\%$ of the maximum current is achieved as a function of the coupling strength $\lambda$ in the case of a fully symmetric junction ($V_L=-V_R=V$, $\Gamma_L=\Gamma_R=0.001\omega_c$ and $k_BT=0.01\omega_c$).
	In dashed black, expected classical value $V_{90}=2\omega_c\lambda^2$ at which the junction gets unblocked.
	}
	\label{fig:Vlim}
\end{figure}

\subsection{Rate equation for $P_n(t)$}

We derive in this section the rate equation for the photon population $P_n(t)=\sum_{q=0,1} \Pi_{qn}(t)$, with $\Pi_{qn}$ the probability of occupation of the state with $q=0,1$ charge on the dot and $n \in \mathbb{N}$ photons of the cavity mode.
We obtain the following rate equation for $\Pi_{qn}(t)$
\begin{eqnarray}
\dot{\Pi}_{qn} &=& \sum_{m=0}^{\infty} \left\lbrack
\Gamma_{(\bar{q}m)(qn)}\Pi_{\bar{q}m} -\Gamma_{(qn)(\bar{q}m)}\Pi_{qn}
\right\rbrack
\nonumber\\
&&+
n\kappa_{\uparrow}\Pi_{qn-1} + \left(n+1\right)\kappa_{\downarrow}\Pi_{qn+1}\nonumber\\
&&-
\left\lbrack
\left(n+1\right)\kappa_{\uparrow} + n\kappa_{\downarrow}
\right\rbrack
\Pi_{qn}  \, ,
\label{eq:RE1} 
\end{eqnarray}
with $\bar{q}=1,0$ when $q=0,1$, $\kappa_{\uparrow}=\kappa n_B(\omega_c)$, and
$\kappa_{\downarrow}=\kappa [1+n_B(\omega_c)]$.
The transition rates $\Gamma_{(qn)(\bar{q}m)}$ involving photon emission or absorption mediated by electron tunneling events are given by Fermi golden rule 
\begin{eqnarray}
\Gamma_{(0n)(1m)} &=& \Gamma |M_{mn}|^2 \sum_\alpha n_F \left(
\left( m - n \right) \omega_c - V_\alpha
\right)
\label{eq:RE2} \\
\Gamma_{(1n)(0m)} &=& \Gamma |M_{nm}|^2 \sum_\alpha \left\lbrack
1 - n_F \left(
\left( n - m \right) \omega_c - V_\alpha
\right),
\right\rbrack\nonumber\\
\label{eq:RE3} 
\end{eqnarray}
and $|M_{mn}|^2$ is the Franck-Condon matrix element \cite{Kovo.05.Franck-Condonblockadeand} for the considered transition.
The stationary solution $P^{st}_n=\lim_{t\rightarrow +\infty}P_n(t)$
of \Eqref{eq:RE1} enables to calculate 
the second order correlation function of the cavity mode at equal time 
\begin{equation}
g^{(2)}(0)=  \frac{\langle n^2 \rangle - \langle n \rangle}{\langle n \rangle^2}
\ ,
\label{G21}
\end{equation}
with $\langle n^k \rangle \equiv \sum_{n=0}^{+\infty} P^{st}_n n^k$.
In the following, we focus on the case of symmetric bias voltage 
$V_L=-V_R=V/2$, such that 
$\Gamma_{(0n)(1m)}=\Gamma_{(1n)(0m)}\equiv \Gamma_{nm}$.
A simpler and closed rate equation can be derived in this case for the photon population $P_n(t)$, after integrating out the charge degree of freedom of the dot
\begin{eqnarray}\label{eq:rate_lim}
\dot{P}_{n} &=& \sum_{m=0}^{\infty} \left\lbrack
\Gamma_{mn} P_{m} - \Gamma_{nm} P_{n}
\right\rbrack
\nonumber \\
&&+
n\kappa_{\uparrow} P_{n-1} + \left(n+1\right)\kappa_{\downarrow}P_{n+1}\nonumber\\
&&-
\left\lbrack
\left(n+1\right)\kappa_{\uparrow} + n\kappa_{\downarrow}
\right\rbrack
P_{n} .
\label{eq:RE4} 
\end{eqnarray}

\subsection{Large bias voltage limit $V_L \ge 2\omega_c$}
\begin{figure}[htb]
	\includegraphics[width=\columnwidth,trim=0 10 5 10,clip]{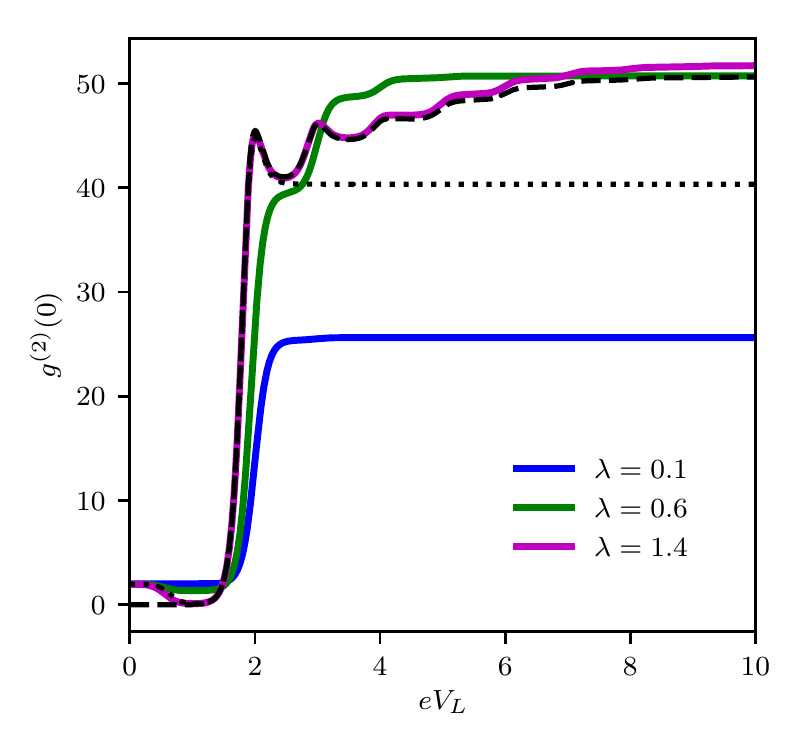}
	\caption{
	Degree of coherence $g^{(2)}(0)$ at time $t=0$ as a function of the left voltage drop $eV_L$ for three different values of the coupling strength $\lambda$. 
	For all three curves, $k_BT=\kappa=0.1\omega_c$ and $\Gamma_L=\Gamma_R=0.001\omega_c$. 
	The black dotted curve shows the analytical prediction at low bias voltage and the black dashed line shows the analytical prediction at large bias voltage for $\lambda=1.4$ (magenta).
	}
	\label{fig:full-bias1}
\end{figure}
From now, we focus on the regime $\kappa \gg \Gamma$, for which the cavity damping rate dominates over the dissipation rate induced by tunneling electrons.
We also consider low-temperatures $T =0.1 \omega_c$, such that $\kappa_{\uparrow} \ll \kappa_{\downarrow} \approx \kappa$.
In this regime, the cavity is thus always close to its quantum ground state, namely $P_0 \approx 1 \gg P_{n \ge 1}$.
For $V_L \ge 2\omega_c$, we keep in \Eqref{eq:RE4} the leading orders in the 
photon populations and rates, such that
\begin{eqnarray}\label{eq:rate_lim}
\dot{P}_{n} &\approx& \Gamma_{0n} P_{0}
+
\kappa \left\lbrack (n+1) P_{n+1} - n P_{n} \right\rbrack \, .
\label{eq:RE5} 
\end{eqnarray}
\Eqref{eq:RE5} can be solved exactly leading to the stationary population
\begin{eqnarray}
P^{st}_{n\ge 1} &=& \frac{P^{st}_0}{\kappa n}\sum_{k\geq n}\Gamma_{0k} \, .
\label{eq:Pn1} 
\end{eqnarray}
Incorporating \Eqref{eq:Pn1} into \Eqref{G21} enables to derive an analytical expression for $g^{(2)}(0)$ that is valid at finite voltage, down to the two-photon emission threshold $V_L \ge 2\omega_c$.
At the second inelastic threshold $V_L \approx 2\omega_c$, emission of photon pairs at equal time is strongly enhanced, thus resulting into a strong bunching of the $g^{(2)}(0)$ versus $V$ curve ( see black dashed curve in Fig.\ref{fig:full-bias1} ).
The infinite bias voltage limit $V \gg \omega_c$ simplifies further in \Eqref{eq:Pn1} to 
\begin{equation}
P^{st}_{n\ge 1}\approx P^{st}_0 \frac{\Gamma}{\kappa n}
\label{eq:Pn2} \, ,
\end{equation}
such that 
\begin{eqnarray}
g^{(2)}(0) \approx \frac{\kappa}{2\Gamma} \, .
\label{eq:Pn2} 
\end{eqnarray}
We thus predict a saturation of $g^{(2)}(0)$ at large voltages to a  
value $g^{(2)}(0) \gg 1$, corresponding to strong bunching of the emitted light. 
For the parameters of Fig.3 in the paper, $\Gamma \approx 10^{-3}\omega_c$ and $\kappa \approx 0.1 \omega_c$, such that $g^{(2)}(0)\approx 50$, as observed in Fig.\ref{fig:full-bias1}.

\subsection{Low-voltage limit $V_L \leq 2\omega_c$}

In the range of voltages $V_L \leq 2\omega_c$, we have to be more careful about how to keep the leading terms in \Eqref{eq:RE4}, in particular in order that 
$P^{st}_{n}$ converges to thermal equilibrium at $V_L=0$.
We found that below the two-photon inelastic threshold ($V_L \leq 2\omega_c$), it is sufficient to keep only states involving at most $2$ photons, such that \Eqref{eq:RE4} can be truncated and simplified as
\begin{eqnarray}
\dot{P}_2 &\approx& \Gamma_{02} P_0 + \left( 2 \kappa_{\uparrow} + \Gamma_{12} \right) P_1 - 2 \kappa_{\downarrow} P_2
\label{eq:PnVlow1} \\
\dot{P}_0 &\approx& \kappa_{\downarrow} P_1 - \left( \kappa_{\uparrow} + \Gamma_{01} + \Gamma_{02} \right) P_0
\label{eq:PnVlow2} \\
P_1 &=& 1 - P_0 - P_2 \, .
\end{eqnarray}
The stationary solution of this system of equations provides 
\begin{equation}\label{g2}
g^{(2)}(0) \approx \frac{\Delta\left\lbrack
\kappa_{\downarrow} \Gamma_{02} + \left(2\kappa_{\uparrow} +\Gamma_{12}\right)\left(\kappa_{\uparrow} +\Gamma_{01}+\Gamma_{02}\right)
\right\rbrack
}{2 \left\lbrack
\left(\kappa_{\uparrow} +\Gamma_{01}+\Gamma_{02}\right)
\left(\kappa_{\downarrow} + 2\kappa_{\uparrow} + \Gamma_{12}\right)
+
\kappa_{\downarrow}\Gamma_{02}
\right\rbrack^2 \,
}    
\end{equation}
with 
\begin{eqnarray}
\Gamma_{0n} &=& e^{-\lambda^2} \frac{\lambda^{2n}}{n!}\sum_\alpha \Gamma_\alpha n_F \left( n\omega_c - V_\alpha \right) \, ,
\label{Rates1}
\end{eqnarray}
the rate for the transition from 0 to n photons in the cavity, 
\begin{eqnarray}
\Gamma_{12} &=& e^{-\lambda^2} \lambda^{2} \frac{\left( 
2 - \lambda^2 \right)^2}{2}\sum_\alpha \Gamma_\alpha n_F \left(\omega_c - V_\alpha \right)
\label{Rates2} \, ,
\end{eqnarray}
the corresponding rate for the transition from 1 to 2 photons in the cavity, 
and $\Delta = \left(\kappa_{\uparrow} +\Gamma_{01}+\Gamma_{02}\right)
\left\lbrack
2\left( \kappa_{\downarrow} + \kappa_{\uparrow}\right) + \Gamma_{12}
\right\rbrack
+
\kappa_{\downarrow}\left(2\kappa_{\downarrow} + \Gamma_{02}\right)
$.
Close to the one-photon inelastic threshold ($V_L \approx \omega_c$), we find that \Eqref{g2} predicts a minimum value for $g^{(2)}(0)$ given by 
$\mbox{ min }g^{(2)}(0)\approx \Gamma_{12}/\Gamma_{01} \approx \left( 2 - \lambda^2 \right)^2/2$.
\Eqref{g2} compares well with the numerics as shown in Fig.\ref{fig:full-bias1} (see black dotted curve).
%

\section{Time dependence of $g^{(2)}(t)$}

\begin{figure}[htb]
	\includegraphics[width=\columnwidth,trim=0 10 10 10,clip]{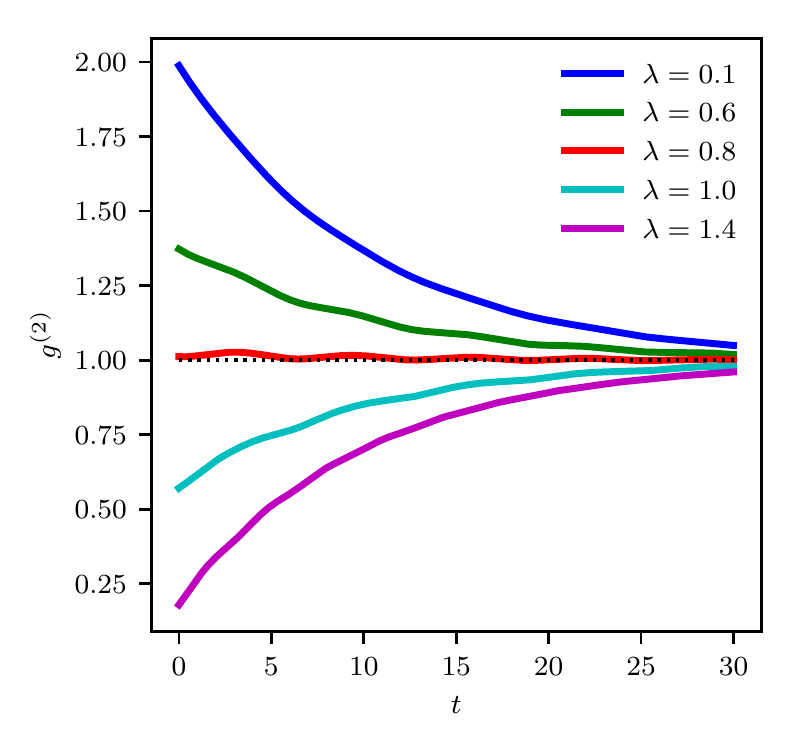}
	\caption{
	Degree of coherence $g^{(2)}(t)$ as a function of time $t$ for three different values of the coupling strength $\lambda$. For all three curves, $k_BT=\kappa=0.1\omega_c$, $\Gamma_L=\Gamma_R=0.001\omega_c$ and $V_L=-V_R=1.2\omega_c$. The black dotted line is at $g^{(2)}=1$.
	}
	\label{fig:time}
\end{figure}
In this section, we are interested in the full time dependence of the second order correlation function $g^{(2)}(t)$ of the plasmonic field
\begin{equation}
g^{(2)}(t) = \frac{\left\langle
a^\dagger a^\dagger(t)a(t)a 
\right\rangle}{ \left\langle
a^\dagger a \right\rangle^2} \equiv
\frac{\left\langle
a^{\dagger2} e^{\mathcal{L}t} a^2 
\right\rangle}{ \left\langle
a^\dagger a \right\rangle^2} \ .
\label{g22}
\end{equation}
$g^{(2)}(t)$ is then computed numerically using the second part of the equation, that is an expression of quantum regression theorem \cite{cohen1998atom}.
The curves $g^{(2)}(t)$ are shown in \Figref{fig:time} for various values of $\lambda$, at a fixed value of bias voltage close to the first photon-emission threshold $V_L=-V_R=1.2 \omega_c$.
In the limit of vanishing plasmon-molecule coupling 
$\lambda \rightarrow 0$, the plasmon mode gets to thermal equilibrium with 
the external electromagnetic bath. 
For a damped electromagnetic cavity at thermal equilibrium, we obtain
\begin{equation}
g^{(2)}(t) = \frac{ \langle a^{\dag2} a^2\rangle e^{-\kappa t}+n_B(\omega_c)\langle a^\dag a\rangle[1-e^{-\kappa t}]}{\langle a^\dag a\rangle^2} \ ,
\label{g2damped}
\end{equation}
with an exponential decay given by $\kappa$ the damping rate of the cavity.
The curve $g^{(2)}(t)$ decays from $g^{(2)}(0)=2$ its value at thermal equilibrium, to a value $g^{(2)}(t\rightarrow+\infty)=1$ characterizing uncorrelated photon emission events at large times.
Note that in this case, since $g^{(2)}(t)<g^{(2)}(0)$ close to $t=0$, the probability of emission of two photons at the same initial time is higher than the one of emitting two successive photons, thus characterizing bunching of the emitted light. 
The fact that $g^{(2)}(t) > 1$ for all times means that the statistics of emitted light is super-Poissonian.
As expected, the obtained blue curve ($\lambda=0.1)$ in \Figref{fig:time} is close to the one predicted by \Eqref{g2damped}.
Upon increasing $\lambda$, however, we notice a crossover towards a different regime
(see magenta curve for $\lambda=1.4$ in \Figref{fig:time}) for which, $g^{(2)}(t)>g^{(2)}(0)$ close to $t=0$, namely the probability of emitting two photons at successive times is higher than the one of emitting two photons at the same time. 
\textit{This unambiguously characterizes photon antibunching of the emitted light \cite{PhysRevA.41.475}.}
Moreover, upon decreasing temperature, antibunching gets almost perfect $g^{(2)}(0)\approx 0 <1$.
The fact that $g^{(2)}(t) < 1$ at any times is a signature that the emitted light has a sub-Poissonian statistics. 
In the intermediate coupling regime $\lambda\approx 0.8$ (red curve in \Figref{fig:time}), $g^{(2)}(t)$ oscillates weakly in time around 1.
The statistics of emitted light is thus close to the Poisson distribution obtained for a coherent classic field.
We note that after checking the behavior of the full time dependence encoded in $g^{(2)}(t)$ (see Fig.\ref{fig:time}), the knowledge of $g^{(2)}(0)$ is sufficient
in Fig.\ref{fig:time} to characterize the existence of a crossover from a regime of bunching ($g^{(2)}(0)>1$) to antibunching ($g^{(2)}(0)<1$) of photon-emission. 

\bibliographystyle{apsrev4-1}
\bibliography{references.bib}